\documentclass[reprint,
nofootinbib,superscriptaddress,amsmath,amssymb,aps,prd]{revtex4-2}
\usepackage[utf8]{inputenc}
\usepackage{graphicx}
\usepackage{hyperref}
\usepackage{url}
\usepackage{color}
\usepackage{wrapfig}
\usepackage{isotope}

\def\be{\begin{equation}}
\def\ee{\end{equation}}
\def\bea{\begin{eqnarray}}
\def\eea{\end{eqnarray}}

\newcommand{\cevns}{CE$\nu$NS}
\newcommand{\xst}{$X_{17}$}

\begin{document}

\title{\textbf{\large $X_{17}$ Existence Hinted at by Nuclear Reactor Neutrinos}}

\author{J.~Rathsman}
\email{Correspondong author: johan.rathsman@fysik.lu.se}
\affiliation{Department of Physics, Lund University,  221 00 Lund, Sweden.}

\author{J.~Cederk{\"a}ll}
\email{joakim.cederkall@fysik.lu.se}
\affiliation{Department of Physics, Lund University, 221 00 Lund, Sweden.}

\author{Y.~Hi\c{c}y{\i}lmaz}
\email{yasarhicyilmaz@balikesir.edu.tr}
\email{yasar.hicyilmaz@physics.uu.se}~
\email{y.hicyilmaz@soton.ac.uk}
\affiliation{Department of Physics, Bal{\i}kesir University,   TR10145, Bal{\i}kesir, Turkey.}
 \affiliation{Department of Physics and Astronomy, Uppsala University, 751 20, Uppsala, Sweden.} 
 \affiliation{School of Physics and Astronomy, University of Southampton, Highfield, Southampton SO17 1BJ, United Kingdom.} 

\author{E.~Lytken}
\email{else.lytken@fysik.lu.se}
\affiliation{Department of Physics, Lund University,  221 00 Lund, Sweden.}

\author{S.~Moretti}
\email{stefano.moretti@physics.uu.se}
\email{s.moretti@soton.ac.uk}
\affiliation{Department of Physics and Astronomy, Uppsala University,  751 20, Uppsala, Sweden.}
\affiliation{School of Physics and Astronomy, University of Southampton, Highfield, Southampton SO17 1BJ, United Kingdom.}%Lines break 

\begin{abstract}
We show that by exploiting 
 the process of Coherent Elastic neutrino ($\nu$) Nucleus Scattering (\cevns), neutrino measurements by nuclear reactor experiments appear to corroborate the evidence of the so-called \xst\ particle, 
which has been invoked to explain the ATOMKI anomaly. We base our analysis primarily on CONUS+ and Dresden-II data, which, when combined with \cevns\ data from COHERENT and neutrino oscillation data from IceCube, single out a unique region of couplings to neutrinos and nuclei.
\end{abstract} 
\maketitle

\textit{Introduction --}
The existence of Coherent Elastic neutrino ($\nu$) Nucleus Scattering (\cevns) was observed experimentally in 2017 by the COHERENT collaboration~\cite{COHERENT:2017ipa}, using secondary neutrinos produced by the Spallation Neutron Source (SNS) at the Oak Ridge National Laboratory (ORNL). The first measurement of \cevns\ using reactor anti-neutrinos  was later performed at the Dresden-II reactor~\cite{Colaresi:2022obx}, Morris, Il, USA and subsequently by the CONUS+ collaboration~\cite{Ackermann:2025obx} at the Leibstadt reactor, Switzerland. The difference between the two approaches is that in the former case higher energy (tens of MeV) neutrinos are produced from $\pi^+$ decay in a spallation target resulting in a pulsed multi-flavor neutrino flux, while in the latter case, low-energy (a few MeV) electron anti-neutrinos are produced in a fission reactor following $\beta^-$ decay. 

In addition to the high-precision measurements needed to detect \cevns, which in itself is a Standard Model (SM) phenomenon, both types of neutrino beams can be used to explore models for Beyond SM (BSM) physics~\cite{Lindner:2016wff}. In particular, this involves the possible existence of new light gauge bosons that e.g.~emerge from $U(1)^\prime$ theories, where the new gauge group can be spontaneously broken by a new gauge boson field (a $Z'$), with a vacuum expectation value in the multi-GeV range. Amongst such $U(1)'$ frameworks, one can use \cevns\ to test the possibility that this $Z'$ is the so-called \xst\ particle that has been invoked to explain the ATOMKI anomaly~\cite{Krasznahorkay:2015iga,
Krasznahorkay:2021joi,Krasznahorkay:2022pxs,Krasznahorky:2024adr}.
The ATOMKI experiment at the Hungarian Institute for Nuclear Research (HUN-REN) in Debrecen~\cite{Gulyas:2015mia} has, over the past few years, studied the properties of $e^+e^-$ pairs generated in nuclear transitions in \isotope[8]{Be}, \isotope[4]{He} and \isotope[12]{C} nuclei 
\cite{Krasznahorkay:2015iga,
Krasznahorkay:2021joi,Krasznahorkay:2022pxs,Krasznahorky:2024adr}.
An excess in both the invariant mass and opening angle of the 
emerging electron-positron pair was consistently seen for these cases with a combined statistical significance well above the 5$\sigma$ level, which points to the possible existence of a new particle named \xst, as the best-fit mass is approximately 17~MeV.  
Furthermore, the PADME collider experiment 
at the Laboratori Nazionali di Frascati (LNF) has also showed sensitivity to this mass range and has presented results in Ref.~\cite{PADME:2025dla}, that show a $\sim$$2\sigma$  excess corresponding to the mass indicated by the ATOMKI experiments. In contrast, the MEG-II experiment at the Paul Scherrer Institut (PSI) in Villingen disfavors the \xst\ hypothesis and has set an upper limit on the $e^+e^-$ decay rate of such a possible new particle~\cite{MEGII:2024urz}. Other facilities also have sensitivity to a possible \xst\ state, including those at 
    CERN (the NA62, NA64 and nTOF experiments), PSI (the Mu3e experiment), 
    JLab (the PRad, DarkLight and HPS experiment) and 
    GANIL (the New JEDI project). Still, none of the experiments that are in operation has reported conclusive
    arguments in either direction~\cite{Alves:2023ree}.
    
As pointed out in Ref.~\cite{Li:2025pfw}, there is currently some tension between the data from the Dresden-II and CONUS+ experiments when analysed within the SM. 
Taking this as a starting point we show in this paper that the two results can be reconciled if the data is analysed in a framework that includes a possible \xst\ particle (for more information about the framework and a more extensive analysis see Ref.~\cite{Rathsman:2026smv}). We find that this analysis is of interest to expand on previous attempts to interpret the ATOMKI results in terms of background nuclear physics or QCD effects~\cite{Zhang:2017zap,Koch:2020ouk,Chen:2020arr,Aleksejevs:2021zjw,Kubarovsky:2022zxm,Hayes:2021hin,Viviani:2021stx}. In addition to reactor neutrino data, our analysis includes data from the COHERENT (see Refs.~\cite{COHERENT:2020iec,COHERENT:2021xmm,COHERENT:2024axu})\footnote{We also note that the European Spallation Source (ESS) in Lund will soon be in operation, and provide further access to \cevns\ processes, which we have shown are sensitive to the existence of the $X_{17}$~\cite{Cederkall:2025bka}} and IceCube \cite{IceCubeCollaboration:2021euf} experiments. For other BSM analyses using nuclear reactor neutrinos we refer to Refs.~\cite{AtzoriCorona:2022qrf,Chattaraj:2025fvx, DeRomeri:2025csu,AtzoriCorona:2025ygn,AtzoriCorona:2025xgj}.

\textit{A Simplified Model for the $Z'$ as \xst\ --}
A minimal theoretical approach that embeds the \xst\ is one where such a state is a $Z^\prime$ gauge boson with interactions with SM fermions given by
\begin{equation}
\label{eq:NeuCurLag}
\mathcal{L}^\mathrm{Z'} = \sum_f \bar f \gamma^\mu \left( C_{f,V}  + \gamma^5 C_{f,A}  \right) f Z'_\mu .
\end{equation}
Here the vector and axial couplings, $C_{f, V}$ and 
$C_{f, A}$, respectively, are taken as arbitrary 
with $f=e, n, p$ and $\nu_{l}$ referring to electrons, neutrons, protons, and neutrinos (letting  $l=e, \mu$ and $\tau$, respectively). This allows for both vector and axial-vector couplings, which are potentially required to explain the ATOMKI anomaly. Given that the neutrinos of the standard model are left handed $C_{\nu_l, V}=C_{\nu_l, A}$, which we write as $C_{\nu_{l}}$ in the following. We also assume that $C_{e, A}=0$  so that the dominant\footnote{There is also a constraint $|C_{p,A} (ZC_{p,V}+NC_{n,V})/A|<6.0\times 10^{-8}$ from \isotope[133]{Cs}~\cite{Dzuba:2017puc} which we also make use of below.} atomic parity violation constraints~\cite{Kahn:2016vjr,Porsev:2009pr,Arcadi:2019uif} are automatically fulfilled. In addition the possible values of $C_{f, V}$ and $C_{f, A}$ can be restricted by applying a suite of experimental constraints provided by the NA64~\cite{NA64:2019auh,NA64:2023wbi}, KLOE~\cite{Anastasi:2015qla}, NA48/2~\cite{NA482:2015wmo}, TEXONO~\cite{TEXONO:2009knm}, and IceCube~\cite{IceCubeCollaboration:2021euf} experiments. 

{\it \cevns\ as a probe for the \xst --}
In order to provide a theoretical prediction, to compare to the experimental spectra, we follow the general framework given in Ref.~\cite{Cederkall:2025bka}. 
Using $y=E_r/E_r^{\max}$ for the fractional nuclear recoil energy and $x=E_\nu/E_\nu^{\max}$ for the fractional neutrino energy, the theoretical nuclear recoil spectrum can be written as
\begin{eqnarray}
\label{eq_dndy}
 \dfrac{d N_r}{dy} & = & m_{\rm det} t_{\rm exp}\dfrac{N_{\rm A} }{M_{\rm A}}    \int_{\sqrt{y}}^{1}  \dfrac{d \sigma^{\nu_eN}}{dy}  \dfrac{d \Phi_{\overline{\nu}_e}}{dx} dx, 
\end{eqnarray}
where $m_{\rm det}$ is the mass of the detector, $t_{\rm exp}$ is the exposure time, $M_{\rm A}$ is the mass of the detector nucleus (see the End Matter), $ N_{\rm A}$ is Avogadro's number, $\frac{d\sigma^{\nu_eN}}{dy}$ is the differential cross-section for CE$\nu$NS and $ \frac{d \Phi_{\overline{\nu}_e}}{dx}$ is the neutrino flux from the reactor. In order to use the same formalism for reactor neutrinos and $\pi^+$ decays at rest, we use $E_\nu^{\max}=m_\mu/2$ and $E_r^{\max}=m_\mu^2/(2M)$. For a Germanium detector, as used in the CONUS+ and Dresden-II experiments, with mass, $M=67.6$ GeV, we have $E_r^{\max}=82.5$~keV. 
The \cevns\ cross-section for the muon and electron neutrino flavors, including an additional $Z^\prime$ boson with mass $m_{Z^\prime}$, can be written as
\begin{eqnarray}
\label{eq_dsigdy}
\dfrac{d\sigma^{\nu_{e/\mu}N}}{dy} & = & \dfrac{ [ N - (1-4\sin^2\theta_W)Z]^2 \, [F_V(y)]^2}{4\pi m_\mu^2}  \, \\ \nonumber && \times \left(1- \dfrac{y}{x^2} \right)
\left(  \dfrac{G_Fm_\mu^2}{\sqrt{2} }  - \dfrac{C^{\nu_{e/\mu}}_{\rm eff}  }{ y + m_{Z^\prime}^2/m_\mu^2}   \right)^2,
\end{eqnarray}
where $\theta_W$ is the weak mixing angle ($\sin^2\theta_W = 0.24$). $C^{\nu_{e/\mu}}_{\rm eff}$ denotes the effective $\nu_{e/\mu}N$ coupling mediated by the $Z^\prime$ particle, given by the vector couplings of the $Z^\prime$ particle to neutrinos, protons and neutrons according to
\begin{eqnarray}
\label{eq_EffectiveNeutrinoCoupling}
C^{\nu_{e/\mu}}_{\rm eff} &=&\dfrac{C_{{\nu_{e/\mu}}} (ZC_{p,V} + N C_{n,V}) }{N - (1-4\sin^2\theta_W)Z} \, .
\end{eqnarray}
$F_V(y)$ is the nuclear form factor for which we use the Klein-Nystrand model~\cite{Klein:1999qj} with  $R_A =5.01 $ fm  and $a=0.7$ fm. Furthermore, we parametrise the neutrino flux from a reactor as described in the End Matter.

In the final step, the predicted spectrum is obtained from the theoretical recoil spectrum Eq.~(\ref{eq_dndy}) by applying the Quenching Factor (QF), which describes the conversion of the nuclear recoil energy to an ionisation signal, and the finite detector resolution giving
\begin{equation}
    \dfrac{dN_r}{dy_{\rm r}} =  \int_{y_{\rm min}}^1 R\left(y_{\rm r},y_{\rm i}\right) \dfrac{dN_r}{dy}  dy \, ,
\end{equation}
where $y_{\rm r}$ is the reconstructed energy fraction and $y_{\rm i}= Q(y)y$ is the fractional ionisation energy.
$Q(y)$ is the QF for which we use
the Lindhard model with  $k = 0.157$. For later comparison we denote this quenching factor as ${\rm QF}_1$.  
In addition we consider a modified QF, which we denote ${\rm QF}_2$,  
to describe the data from Ref.~\cite{Collar:2021fcl}. We discuss this QF in more detail in~\cite{Rathsman:2026smv}. To quantify the difference between the quenching factors, we also define a QF uncertainty,
\begin{equation}
\label{eq_QFuncertainty}
 \Delta_{\rm QF} = \left| \left.\dfrac{dN_r}{dy_{\rm r}}\right|_{\rm QF2} -\left.\dfrac{dN_r}{dy_{\rm r}}\right|_{\rm QF1}  \right|,
\end{equation}
as described in more detail in the same paper.
The integration limit, $y_{\rm min}$, is the minimal fractional 
energy needed to create an electron-hole pair in Germanium. The detector resolution is in turn given by 
\begin{equation}
R(y_{\rm r},y_{\rm i})=\dfrac{2}{1+ {\rm Erf }(\frac{y_{\rm i} }{\sqrt{2} \sigma})} \dfrac{1}{\sqrt{2\pi} \sigma } \, e^{ - \dfrac{(y_{\rm r} - y_{\rm i})^2}{2\sigma^2 }}
\end{equation}
where $\sigma=\sigma_E/E_r^{\max}$ is a dimensionless width with $\sigma_E^2 =  \sigma_n^2 + E_{\rm i} \eta F$,  where $E_{\rm i}$ is the ionisation energy, F the Fano factor and $\sigma_n$ is the intrinsic resolution of the detector (for the numerical values used see the End Matter)

\begin{figure}[th]
\begin{center}
\includegraphics[width=6.5cm]{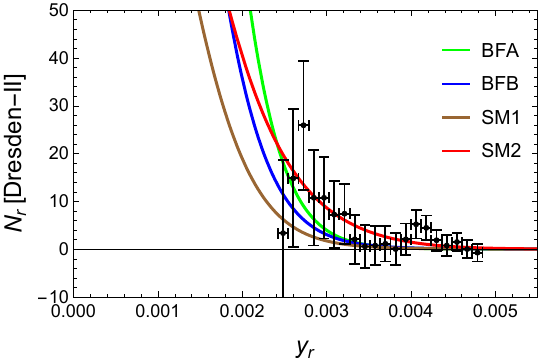} 
\includegraphics[width=6.5cm]{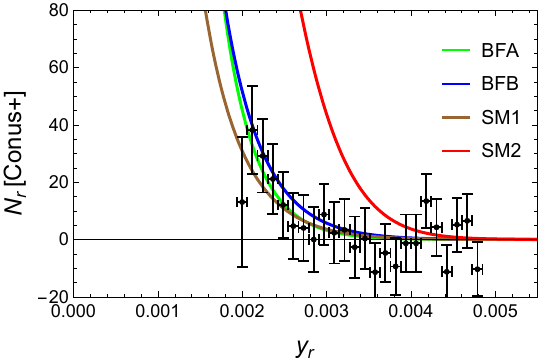} 
\end{center}
\vspace*{-0.3cm}
\caption{Nuclear recoil spectra after smearing and quenching compared to data as published using the QF1 or QF2 in the SM as well as the two best fit points, BFA and BFB, when including a \xst. The data has been normalised in the same way as done by the respective experiments.}
\label{fig_recoilspectrasmeareddata}
\end{figure}
{\it Statistical analysis --}
The measured recoil spectra, presented in Fig.~\ref{fig_recoilspectrasmeareddata}, show that the CONUS+ data are in good agreement with the SM when using ${\rm QF}_1$, but less so using ${\rm QF}_2$ whereas the Dresden-II data is in better agreement with the SM when using ${\rm QF}_2$. In order to quantify these statements further we use the following $\chi^2$ function to provide a quantitative statistical measure:
\begin{equation}
\label{eq_chi2}
\chi^2=\sum_i \left(\dfrac{(1+\rho)x_i-\mu_i}{\sigma_i}\right)^2+ \left(\dfrac{\rho}{\sigma_{\rm sys}}\right)^2 \, ,
\end{equation}
where $x_i$ are the model predictions for the number of events in a bin and $\mu_i$ are the number of observed events with $\sigma_i$ being the corresponding errors. In addition, $\rho$ is an overall scale factor, with a systematic uncertainty $\sigma_{\rm sys}$, that is determined by minimising the $\chi^2$ with respect to $\rho$.
When analysing the two reactor datasets, we will for consistency use the same scaling uncertainty, $\sigma_{\rm sys}=0.17$, for both data sets following Ref.~\cite{Ackermann:2025obx}.
With $\chi^2$ defined as in Eq.~(\ref{eq_chi2}) we get $\chi^2=11.6 \,(37.4)$ for CONUS+ with ${\rm QF}_1$(${\rm QF}_2$) and $\chi^2=14.4 \,  (8.2)$ for Dresden-II with ${\rm QF}_1$(${\rm QF}_2$) in the SM.

\begin{figure}[!t]
\begin{center}
\includegraphics[width=4.2cm]{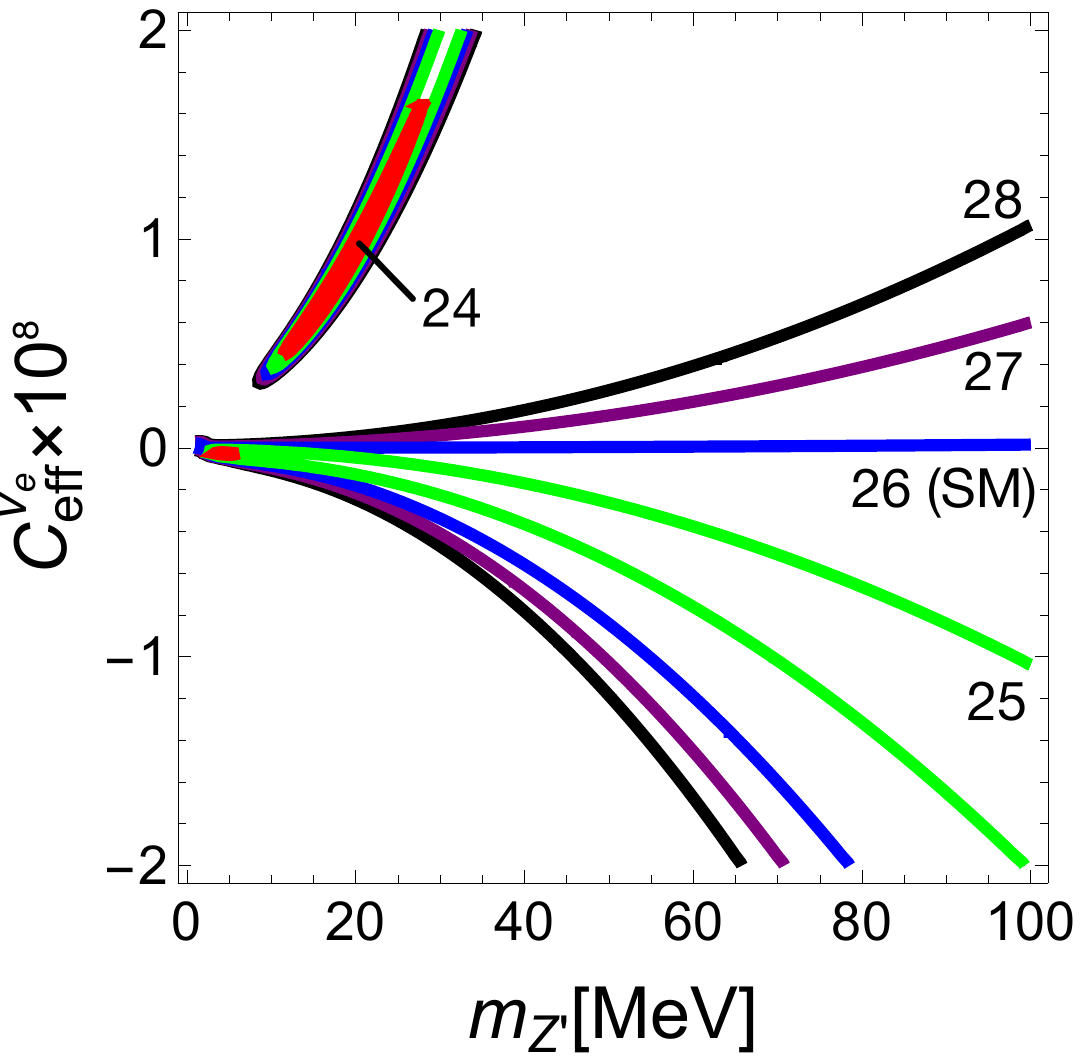} 
\vspace*{0.2cm}
\includegraphics[width=4.2cm]{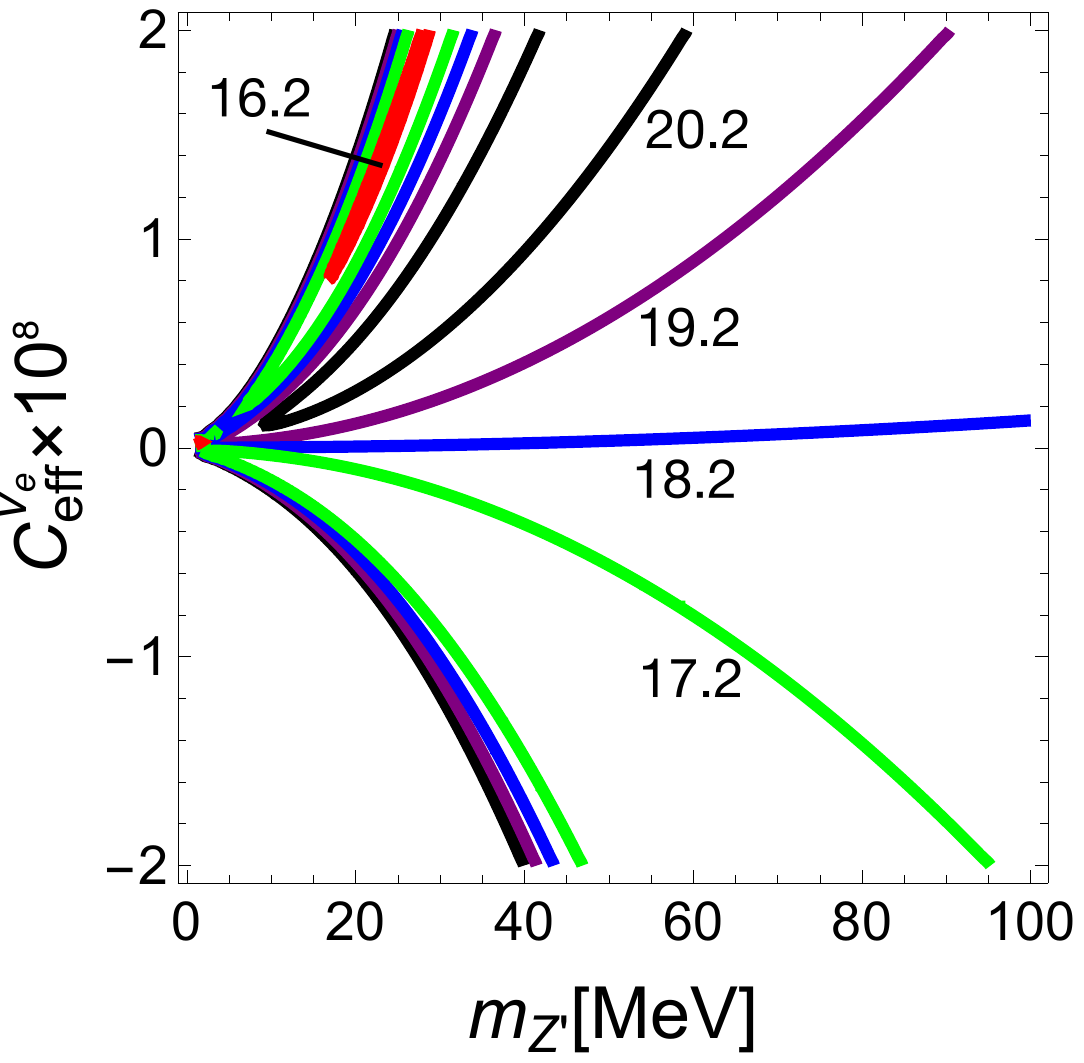} 
\end{center}
\vspace*{-0.5cm}
\caption{The $\chi^2$ contours for reactor datasets using QF1 without (left) and with (right) quenching factor uncertainty added (see text). The SM $\chi^2$ value is given by the sum of the individual $\chi^2$ values for the Dresden-II and Conus+ cases as shown in Fig.~\ref{fig_recoilspectrasmeareddata}. }
\label{fig_chi2ReactorStat}
\end{figure}

As discussed above it is possible to explain the apparent differences between the CONUS+ and Dresden-II experiments by an additional light $Z^\prime$ as defined by the cross-section in Eq.~(\ref{eq_dsigdy}). 
Such an additional light $Z^\prime$ could have any mass and effective coupling to nuclei. We have therefore calculated the $\chi^2$ defined above, when varying these two parameters to investigate what ranges are preferred/allowed by the reactor data. Such an analysis shows (see also paper~\cite{Rathsman:2026smv}) that there is a preferred region $m_{Z^\prime}\approx 20 \pm 10 $ MeV and $C^{\nu_{e}}_{\rm eff} \approx (1 \pm 0.5) \cdot 10^{-8}$, as shown in the left panel of Fig.~\ref{fig_chi2ReactorStat}, which has $\chi^2 \approx 24$ compared to the SM value 26. Adding the quenching factor uncertainty, defined by Eq.~(\ref{eq_QFuncertainty}),
quadratically to the experimental errors bin-by-bin as well as using $\sigma_{\rm sys}=0.15$~\cite{Ackermann:2025obx} since the quenching factor uncertainty is treated separately, the position of the $\chi^2$-minimum shifts slightly as shown in the right part of Fig.~\ref{fig_chi2ReactorStat}, but the overall features are the same. In both cases, we find that including a  light $Z^\prime$ with a mass around 20 MeV improves the description of the data. The apparent differences in how this light $Z^\prime$ affects the recoil spectra for the two experiments, is due to the different detector resolutions of the two as explained in more detail in~\cite{Rathsman:2026smv}.

Before comparing the reactor datasets to a specific \xst\ model, we introduce additional \cevns\ data from the COHERENT experiment, including the argon~\cite{COHERENT:2020iec} and caesium iodine~\cite{COHERENT:2021xmm} targets as well as more recent germanium data~\cite{COHERENT:2024axu}. We use the same framework as above to analyse this data, although it is generalised to include also muon neutrinos. In short, the cross-section for $\nu_{e/\mu}$ nucleus scattering is given by Eq.~(\ref{eq_dsigdy}) and the neutrino fluxes from the $\pi^+$ decay at rest are well known. Convoluting the cross-section with the neutrino fluxes then gives the theoretical nuclear recoil spectrum for neutrinos from pion decays as given in e.g.~Ref.~\cite{Cederkall:2025bka}. For the COHERENT data we use the experimental information (QF, detector resolution and acceptance) as published in Refs.~\cite{COHERENT:2020iec,COHERENT:2021xmm}. In the same way as for the reactor data, we use ${\rm QF}_1$ for the Germanium case. The systematic scale uncertainties used for the different COHERENT datasets are: $\sigma_{\rm sys}^{\rm Ar}=0.13$, $\sigma_{\rm sys}^{\rm CsI}=0.12$, and  $\sigma_{\rm sys}^{\rm Ge}=0.103$. 

The effective neutrino nucleus couplings, given by Eq.~(\ref{eq_EffectiveNeutrinoCoupling}),
include a dependence on the number of neutrons and protons. The main effect is that the $C_{n,V}$ coupling carries a $\sim$20\% larger weight than $C_{p,V}$ from $N$ being larger than $Z$ for the nuclei of interest. In order to only use two effective neutrino nucleus couplings we approximate this dependence by setting $C^{\nu_{e/\mu}}_{\rm eff}({\rm Ar})=1.020 \, C^{\nu_{e/\mu}}_{\rm eff}({\rm Ge})$ and $C^{\nu_{e/\mu}}_{\rm eff}({\rm CsI})=0.953  \, C^{\nu_{e/\mu}}_{\rm eff}({\rm Ge})$, which are obtained by taking the average of the nuclear dependence of the effective neutrino quark couplings defined analogously to Eq.~(\ref{eq_EffectiveNeutrinoCoupling}). 

\begin{figure}[!t]
\begin{center}
\includegraphics[width=4.2cm]{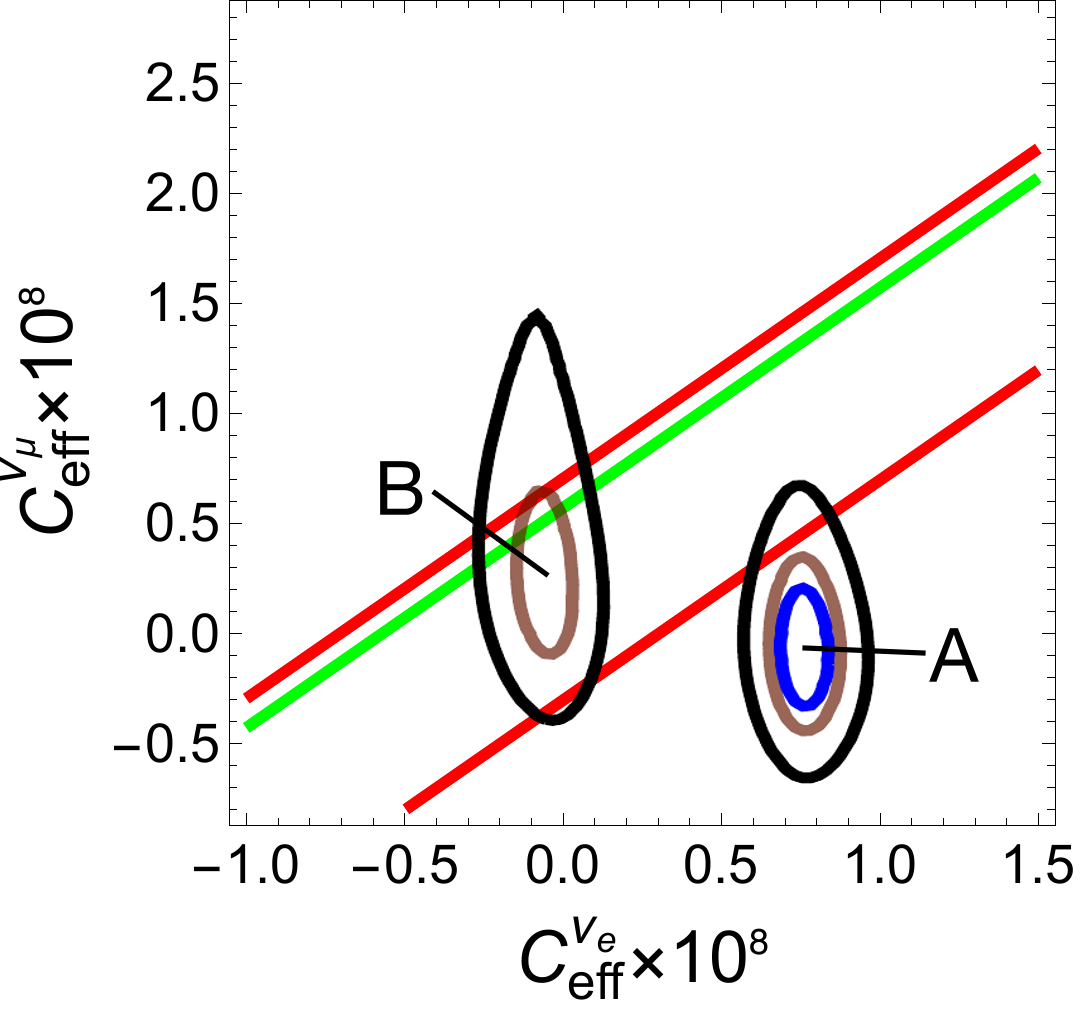} 
\vspace*{0.2cm}
\includegraphics[width=4.2cm]{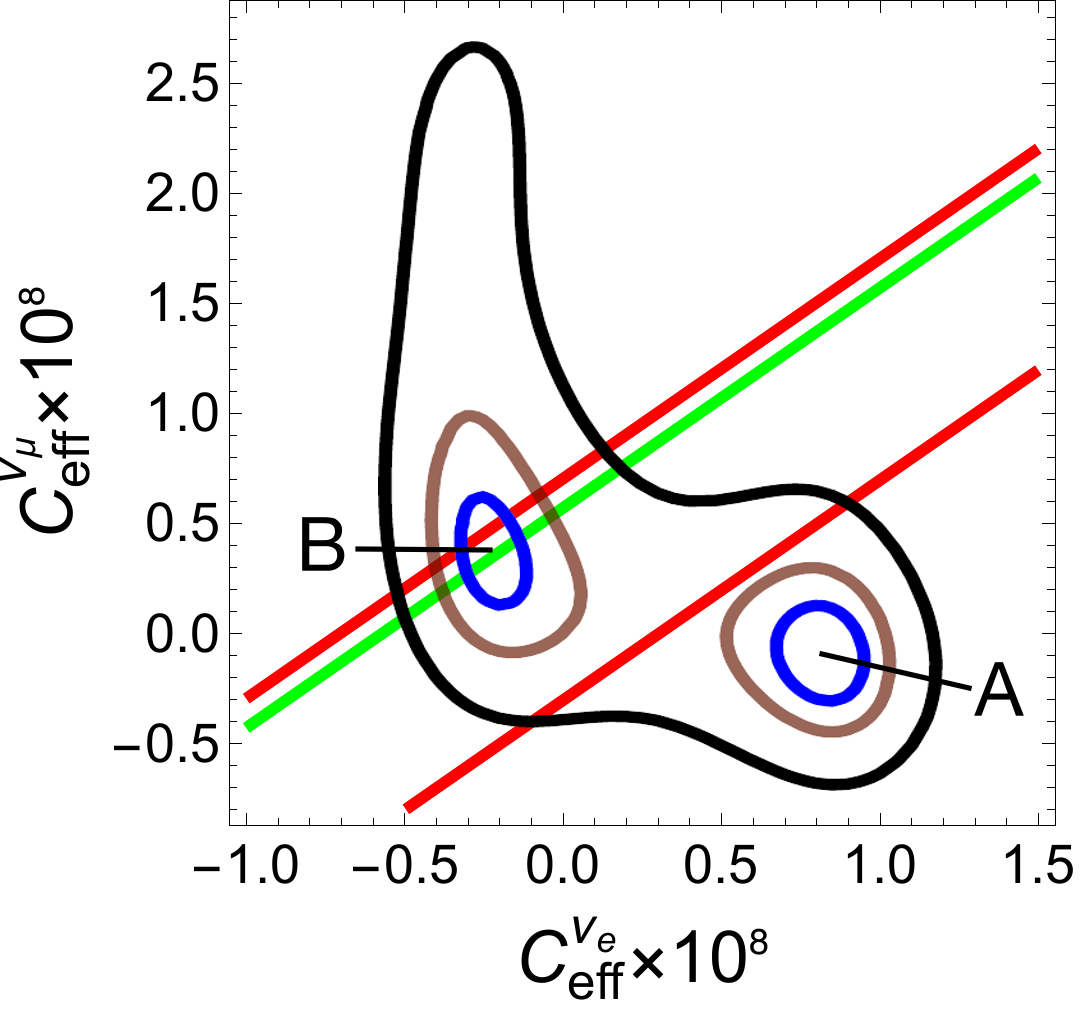} 
\end{center}
\vspace*{-0.5cm}
\caption{The $\chi^2$ contours for  combined reactor and COHERENT datasets  for $m_{Z^\prime}=17$ MeV. Brown denotes $\chi^2=\chi^2_{\rm SM}$, black $\chi^2=\chi^2_{\rm SM}+6$ and red $\chi^2=\chi^2_{\rm SM}-2$  whereas the best-fit regions BFA and BFB are indicated by A and B respectively. Left (right) plot without (with) the additional quenching factor uncertainty added separately. The red lines are the 90\% CL limits from IceCube (with the allowed region being between the two lines) and the green line is their best fit in the case that $C_{n,V}$ dominates both NSI and \cevns.}
\label{fig_chi2AllStat}
\end{figure}

With these definitions, we calculate the predicted recoil spectra also for the COHERENT data sets and compare to the published data using the $\chi^2$ function defined by Eq.~(\ref{eq_chi2}).
The resulting $\chi^2$ for $m_{Z^\prime}=17$ MeV and as a function of the effective neutrino nucleus couplings - $C^{\nu_{e}}_{\rm eff}$ and  $C^{\nu_{\mu}}_{\rm eff}$ - is displayed in Fig.~\ref{fig_chi2AllStat}. It is clear that the COHERENT data does not change the preferred regions for $C^{\nu_{e}}_{\rm eff}$. At the same time, the COHERENT data constrains the $C^{\nu_{\mu}}_{\rm eff}$ coupling, although not quite as strongly as the reactor data constrain $C^{\nu_{e}}_{\rm eff}$.  Our analysis also shows that, although currently the uncertainties from the quenching factor of Germanium are very large for the reactor data, the preferred regions for the effective neutrino-nucleus couplings are more or less the same irrespective of whether this systematics is included or not in our fits~\cite{Rathsman:2026smv}.

Finally, we note that theoretical interpretations of the \xst\ are numerous in literature ~\cite{Feng:2016jff,Feng:2016ysn,Ellwanger:2016wfe,Feng:2020mbt,Nomura:2020kcw,Seto:2020jal,Kozaczuk:2016nma,DelleRose:2018pgm,DelleRose:2019hnc,DiLuzio:2025ojt,Zhang:2020ukq,Barducci:2022lqd,Denton:2023gat,Alves:2023ree,Batra:2026tzz}. From these studies it has been noted that theoretical frameworks embedding a pure vector mediator  are less likely,  while an axial-vector state appears as the best candidate to comply with all current experimental datasets. Consequently, for clarity we point out that in \cevns\ only the vector component of the $X_{17}$ can be tested, but that as we show above it provides one path to reconcile the results of the two reactor based \cevns\ experiment mentioned above. Specifically, it can contribute to the cross-section through both constructive and destructive interference with the $\gamma$ and $Z$ boson of the SM. We also note that, until our recent study~\cite{Cederkall:2025bka}, only flavor universal models for a light $Z^\prime$ in this mass range,  where the couplings to electron and muon neutrinos are the same, had been considered~\cite{Abdullah:2018ykz,Flores:2020lji,Cadeddu:2020nbr,Banerjee:2021laz,AtzoriCorona:2022moj,Coloma:2022avw}.

All in all, from Fig.~\ref{fig_chi2AllStat} we identify two distinct best fit points (or regions):  
\begin{itemize}
    \item BFA: $( C^{\nu_e}_{\rm eff},C^{\nu_\mu}_{\rm eff}) \approx (  0.8  ,  -0.1 )\times 10^{-8}$
    \item BFB: $( C^{\nu_e}_{\rm eff},C^{\nu_\mu}_{\rm eff}) \approx ( -0.2,  0.3) \times 10^{-8}$ 
\end{itemize} 
which more or less fit the \cevns\ data equally well. 

\textit{Analysis using a protophobic \xst\ --}
To analyse the \cevns\ results further one can use the so-called protophobic \xst\ model, where $|C_{p,V}| \ll  |C_{n,V}| $. In addition we assume that $\textrm{BR}(Z^\prime \to \textrm{invis}) \ll \textrm{BR}(Z^\prime \to e^+e^-)$. From the NA48/2 experiment~\cite{NA482:2015wmo} we then have the constraint $|C_{p,V}| < 3.6 \times 10^{-4}$. Furthermore, using the analysis by Barducci and Toni~\cite{Barducci:2022lqd},
$|C_{n,V}| \approx (5\pm2)\times 10^{-3}$ gives a good description of the ATOMKI \isotope[4]{He} and \isotope[8]{Be} data, although there is some tension with the \isotope[12]{C}-data as discussed in~\cite{Fieg:2026zkg}. For these coupling values\footnote{We also note that these couplings gives a constraint $|C_{p,A}|<2\times 10^{-5}$ from atomic parity violation~\cite{Dzuba:2017puc}.} the two best fit regions correspond to 
\begin{itemize}
    \item 
BFA:  $(C_{\nu_e},C_{\nu_\mu}) \approx (1.6\pm0.6,-0.2\pm0.3)\times 10^{-6}$ ,
    \item
BFB: $(C_{\nu_e},C_{\nu_\mu}) \approx (-0.4\pm0.3,0.6\pm0.3)\times 10^{-6}$ .
\end{itemize}
Thus, both BFA and BFB fulfil the NA64 constraint from DM searches~\cite{NA64:2023wbi}, $|C_{e,V}|  {\sqrt{\textrm{BR}(Z^\prime \to \textrm{invis})}} < 1\times 10^{-5}$ which, using
$\textrm{BR}(Z^\prime \to \textrm{invis}) \ll \textrm{BR}(Z^\prime \to e^+e^-)$, can be written as 
\begin{equation}
    \sqrt{ 2( C_{\nu_e}^2  + C_{\nu_\mu}^2+ C_{\nu_\tau}^2 )} < 1\times 10^{-5}.
\end{equation}

In addition, from the NA64~\cite{NA64:2019auh} and KLOE~\cite{Anastasi:2015qla} experiments we have the limits (assuming $|C_{e,A}|=0$ as already stated):  
\begin{equation}
2 \times 10^{-4}<|C_{e,V}|  {\sqrt{\textrm{BR}(Z^\prime \to e^+e^-)}} < 6 \times 10^{-4} ,
\end{equation}
from which it follows that both BFA and BFB fulfil the TEXONO~\cite{TEXONO:2009knm} neutrino-electron scattering constraint, $\sqrt{  |C_{e,V} C_{\nu_e}| } < 9\times 10^{-5}$.
However, the Non-Standard neutrino Interactions (NSI) 90\% CL constraint from IceCube~\cite{IceCubeCollaboration:2021euf},
$-0.74< \epsilon_{ee}^{\oplus }-\epsilon_{\mu\mu}^{\oplus } < 0.32$, 
is not fulfilled by our BFA. 
To clarify this we first rewrite $\epsilon_{ee}^{\oplus }-\epsilon_{\mu\mu}^{\oplus }$ in terms of the Lagrangian in Eq.~(\ref{eq:NeuCurLag}), which gives
\begin{equation}
\epsilon_{ee}^{\oplus }-\epsilon_{\mu\mu}^{\oplus } = \dfrac{C_{\nu_{e}}-C_{\nu_{\mu}}}{2\sqrt{2}G_F m_{Z^\prime}^2} \left(C_{e,V}+C_{p,V}+\dfrac{N_{\oplus }}{Z_{\oplus }}C_{n,V}\right) 
\end{equation}
where ${N_{\oplus }}/{Z_{\oplus }}=1.05$ is the ratio of the number of neutrons to protons in the earth. Secondly, the definition of the effective \cevns\ couplings in Eq.(\ref{eq_EffectiveNeutrinoCoupling}) gives for Ge nuclei 
\begin{equation}
C^{\nu_{e}}_{\rm eff} - C^{\nu_{\mu}}_{\rm eff} = (C_{{\nu_{e}}}-C_{{\nu_{\mu}}})(0.81 C_{p,V} + 1.03C_{n,V}).
\end{equation}
So for the case at hand, with $|C_{p,V}|,|C_{e,V}| \ll |C_{n,V}|$ we have
$C^{\nu_e}_{\rm eff}-C^{\nu_\mu}_{\rm eff}=2\sqrt{2}G_F m_{Z^\prime}^2(\epsilon_{ee}^{\oplus }-\epsilon_{\mu\mu}^{\oplus })$ and we get the NSI derived limit
\begin{equation}
    -0.70\times10^{-8} < C^{\nu_e}_{\rm eff}-C^{\nu_\mu}_{\rm eff} < 0.30\times10^{-8},
\end{equation}
which is also shown in Fig.~\ref{fig_chi2AllStat} together with the NSI best fit point  $\epsilon_{ee}^{\oplus }-\epsilon_{\mu\mu}^{\oplus }=-0.60$. From the figure it is clear that BFA is ruled out by the NSI constraint at 90\% CL, whereas the BFB region on the other hand is in good agreement with the best fit value from NSI.

\textit{Conclusions --}
We have analysed nuclear recoil spectra arising from \cevns\ exploiting electron anti-neutrinos from nuclear reactors in the presence of a light $Z^\prime$ boson. A statistical analysis of the CONUS+ and Dresden-II data, addressing the question of their compatibility with this effect, or the SM hypothesis, indicates that the BSM hypothesis is preferred to the latter. Including this $Z'$ as the \xst, the existing tension between the CONUS+ and Dresden-II data under the SM assumption is resolved, in the sense that it can be explained in terms of their detector resolutions, which in turn give a different sensitivity to the \xst for the two measurements.

Furthermore,  the results indicate that the preferred effective coupling between the electron anti-neutrinos and nuclei mediated by the \xst\ can be both positive or negative. Adding spallation source data from COHERENT, we have illustrated that the effective couplings between muon (anti-)neutrinos and nuclei mediated by the \xst\ prefer two distinct regions. Finally, applying also NSI data from IceCube, we find that a single region of effective neutrino-nucleus couplings is singled out with the preferred effective coupling between the electron anti-neutrinos and nuclei mediated by the \xst\ being negative, whereas that between muon neutrinos and nuclei mediated by the \xst\ is positive. These results, obtained in terms of a simplified model description adding the \xst\ to the SM, can be used as further input to appropriate model building.

In conclusion, we have demonstrated that there are {\sl indications} of the existence of the \xst\ from nuclear reactor, spallation source and NSI data, which is consistent with, yet independent of, a variety of spectrometer data.  

\textit{Acknowledgments --}
We thank Max H Fieg for bringing Ref.~\cite{Dzuba:2017puc} to our attention. The work of YH is supported by The Scientific and Technological Research Council of Turkey (TUBITAK) in the framework of the 2219-International Postdoctoral Research Fellowship Programme. SM is supported in part through the NExT Institute and STFC Consolidated Grant ST/X000583/1. 
YH, SM and JR thank ECT$^*$ for support at the Workshop “The X17 particle, status and new ideas” during which this work was completed.

\section*{END MATTER}
For clarity we provide the numerical values for a number of parameters used in the analysis in the following. We parametrise the reactor neutrino flux using 
\begin{equation}
 \frac{d \Phi_{\overline{\nu}_e}}{dx} = 
 \frac{ \Phi (e^{a_1 + b_1 x + c_1 x^2 + d_1 x^3 + e_1 x^4 + f_1 x^5}+ xe^{a_2 + b_2 x + c_2 x^2} )}{N_{\overline{\nu}_e}}  ,
\end{equation}
where the total fluxes for the Dresden-II and CONUS+ experiments are: $\Phi_{\rm Dresden-II} = 4.8 \times 10^{13}$ cm$^2/$s~\cite{Colaresi:2022obx},
and $\Phi_{\rm CONUS+} = 1.5 \times 10^{13}$ cm$^2/$s~\cite{Ackermann:2025obx} respectively. 
The number of neutrinos per fission is $N_{\overline{\nu}_e}= 6.8$, while the remaining parameters are given by $a_1 = 2.87593$, $b_1 = 186.145$, $c_1 = -3358.36$, $d_1 = 31940.8$, $e_1 = -139054$, $f_1 = 193450$, $a_2 = 11.2864$, $b_2 = -149.915$ and $c_2 = -452.33$.
These parameters were obtained by fitting to the Daya Bay data~\cite{DayaBay:2022eyy,DayaBay:2021dqj} for $x>0.0379$ ($E_\nu > 2$ MeV) and then subsequently to the calculation by Kopeikin~\cite{Kopeikin:2012zz}. 

The following parameters were used for calculating the the nuclear recoil spectrum and the resolution.  The detector masses and exposure times are $m_{\rm det}=2.83$ kg and  $t_{\rm exp}= 115.5$ days for CONUS+ whereas  $m_{\rm det}=2.924$ kg and $t_{\rm exp}= 96.4$ days  for Dresden-II. For the resolution a Fano factor $F=0.11$ was used with $\eta=2.96$ eV$_{\rm ee}$, and $\sigma_n^{\rm Dresden-II} = 68.5 \, {\rm  eV}_{\rm ee}$  and $\sigma_n^{\rm CONUS+} = 20.0  \, {\rm  eV}_{\rm ee}$ for the two experiments.

\bibliography{refs}
\bibliographystyle{apsrev}

\end{document}